\documentclass{llncs}
\usepackage{pgf}
\usepackage{tikz}
\usetikzlibrary{calc}
\usepackage{graphicx}
\usepackage{amsmath}
\usepackage{amsfonts}
\usepackage{latexsym}
\usepackage{amssymb}
\usepackage{epsfig}
\usepackage{framed}
\usepackage[mathcal]{euscript}

\begin{document}

\title{Discrete geodesics and cellular automata}

\author{Pablo Arrighi\inst{1} \and Gilles Dowek\inst{2}}
\institute{
Aix-Marseille Univ., LIF, F-13288 Marseille Cedex 9, France\\
\email{pablo.arrighi@univ-amu.fr}, 
\and
Inria, 23 avenue d'Italie, CS 81321, 75214 Paris Cedex 13, France.\\
\email{gilles.dowek@inria.fr}
}

\maketitle

\begin{abstract}
This paper proposes a dynamical notion of discrete geodesics,
understood as straightest trajectories in discretized curved
spacetime. The notion is generic, as it is formulated in terms of a
general deviation function, but readily specializes to metric spaces
such as discretized pseudo-riemannian manifolds. It is effective:
an algorithm for computing these geodesics naturally follows, which
allows numerical validation---as shown by computing the perihelion
shift of a Mercury-like planet. It is consistent, in the continuum
limit, with the standard notion of timelike geodesics in a
pseudo-riemannian manifold. Whether the algorithm fits within the
framework of cellular automata is discussed at length.\\
{\bf Keywords:} Discrete connection, parallel transport, general
relativity, Regge calculus.
\end{abstract}

\section{Introduction}\label{sec:intro}

Three reasonable hypotheses---bounded velocity of propagation of
information, homogeneity in time and space, and bounded density of
information---lead to the thesis that natural phenomena can be
described and simulated by cellular automata. This implication has in
fact been formalized into a theorem both in the classical \cite{Gandy}
and the quantum case \cite{ArrighiGandy}, albeit in flat
space. Further evaluating this thesis leads to the project of
selecting specific physical phenomena, such as gravitation, and
attempting to describe them as cellular automata. A first step in this
direction is to build discrete models of the phenomena. In the case of
gravitation, this leads to the question we address in this paper: what
is a discrete geodesics?

Geodesics generalize the flat space notion of line, to curved
spaces. A line is both the shortest, and the straightest path between
two points, but in curved space the two criteria do not coincide
\cite{martinez2005computing}. In computer graphics and discrete
geometry, discrete geodesics as shortest path between two given point
have been studied extensively
\cite{mitchell1987discrete,peyre2010geodesic}. This is not the case of
geodesics as straightest path given an initial point and
velocity---with the noticeable exception of
\cite{polthier2006straightest}, in the framework of simplicial
complexes. Yet, it is this criterion that one must adopt in order to
describe and simulate the timelike geodesics trajectories of
particles. In this paper we adopt the dynamical, spacetime view on
geodesics which is typical of numerical relativity
\cite{vincent2012}. But instead of discretizing geodesics defined by 
partial differential equations in a continuous spacetime, we seek to
discrete geodesics as a native notion of a discretized spacetime, for
instance of a grid endowed with a metric.

More precisely, the paper proposes both a notion of discrete-spacetime
\linebreak
geodesics and a notion of discrete-time continuous-space geodesics
(Section \ref{sec:geodesics}). Both are generic, that is formulated in terms
of a general deviation function, but readily specialize for metric
spaces (Section \ref{sec:distance}). They are effective: an algorithm
for computing timelike geodesics naturally follows
(Section \ref{sec:algo}), which allows us to validate the notions
numerically, by computing the perihelion shift of a Mercury-like
planet (Section \ref{sec:planet}). They are consistent with one another:
the former is clearly a discretization the the latter. Moreover, the
latter is proven to have the standard notion of continuous-spacetime
geodesics in a Riemanian space as its limit, which validates both
notions as legitimate discrete counterparts
(Section \ref{sec:continuum}). Whether the algorithm fits within the
framework of cellular automata is discussed at length, as well as how
this impacts on precision
(Sections \ref{sec:CAMechs}-\ref{sec:precision}).

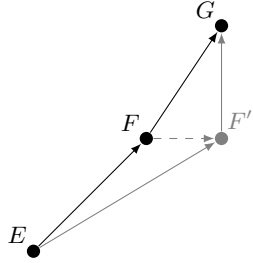
\begin{figure}
\begin{center}
\begin{tikzpicture}
\tikzstyle{every node} = [label distance=1pt, circle, fill, inner sep=0pt, minimum size=5pt]
\tikzset{>=latex}
\node[label=135:$E$] (E) at (-1.5,0) {};
\node[label=135:$F$] (F) at (0,1.5) {};
\node[label=45:{\color{gray}$F'$},gray] (F') at (1,1.5) {};
\node[label=135:$G$] (G) at (1,3) {};
\draw[->] (E) -- (F) node [midway,fill=none,label=45:,label=-93:] {};
\draw[->] (F) -- (G) node [midway,fill=none,label=45:,label=-95:] {};
\draw[->,gray,dashed] (F) -- (F') node [midway,fill=none,label=90:] {};
\draw[->,gray] (E) -- (F') node [midway,fill=none] {};
\draw[->,gray] (F') -- (G) node [midway,fill=none] {};
\end{tikzpicture}
\end{center}
\caption{{\bf Discrete geodesics} seek to find $G$ such that ${\bf FG}$ minimizes its deviation relative to ${\bf EF}$. In the case of metric spaces, ${\bf FG}$ is understood to ``deviate towards ${\bf FF'}$ relative to ${\bf EF}$'', whenever $l(E,F',G) < l(E,F,G)$---such deviations must be
minimized.\label{fig:discretemodel}}
\end{figure}

These results apply to natively discrete formulations
of General Relativity such as Regge calculus
\cite{williams1981regge,brewin1993particle}. For instance, our method
yields perihelion shift computations of the right order, an issue in
\cite{williams1981regge} pointed out in \cite{brewin1993particle}.  We
discuss how our approach differs from \cite{polthier2006straightest}
and why it fixes this issue.  Finally, an often underestimated
contribution is the pedagogical: the simple discrete model summarized
in Figure \ref{fig:discretemodel} has continuum limit the complicated,
well-known equations of \eqref{eq:T2} and \eqref{eq:T3}.

Besides assessing this ``digital physics'' program, we believe that
these results can be applied in any inherently discrete geometrical
setting, in order to compute geodesics without the need to interpolate
a continuous surface. Such applications may arise in computer vision
and graphics \cite{peyre2010geodesic} including computer anatomy
\cite{lorenzi2011schild}.

\section{Discrete geodesics}\label{sec:geodesics}

Consider a discrete-time continuous-space spacetime ${\mathbb Z}
\times {\mathbb R}^n$ where ${\mathbb Z}$ is the discrete timeline and
${\mathbb R}^n$ a continuous space. Consider a {\em deviation}
function $w$ from $({\mathbb Z} \times {\mathbb R}^n)^3$ to ${\mathbb
R}^+$, the number $w(E,F,G)$ measuring how the path $E, F, G$
deviates from ``going straight ahead''. In this setting, a {\em
geodesic} is a sequence of points in ${\mathbb Z} \times {\mathbb
R}^n$ $(E_i)_i$ such that for any $i$, $w(E_{i-1}, E_{i}, E_{i+1}) =
0$.  Such a property can be read as a condition on $E_{i+1}$: if the
points $E_{i-1}$ and $E_i$ are given, the geodesics must continue with
a point $E_{i+1}$ such that $w(E_{i-1}, E_i, E_{i+1})=0$.

Consider now a discrete spacetime $M = {\mathbb Z} \times {\mathbb
  Z}^n$ where ${\mathbb Z}$ is the discrete timeline and ${\mathbb
  Z}^n$ a discrete space and a deviation function $w$ from $M^3$ to
${\mathbb R}^+$ which, as above, measures how the path $E, F, G$
deviates from going straight ahead.  In this setting we cannot demand
$w(E_{i-1}, E_i, E_{i+1})$ to be exactly zero, but we demand that it
be a minimum with respect to {\em spatial local variations} of
$E_{i+1}$.

{\em Spatial local variations} can be defined as follows.  Let us
write $\langle x_0, x_1, ..., x_{n} \rangle$, the coordinates of a
point $E$ in $M$, where $x_0$ is the time coordinate and $x_1,...,
x_n$ the space coordinates.  Two points of $M$, $\langle x_0, x_1,
..., x_n \rangle$ and $\langle x'_0, x'_1, ..., x'_n \rangle$ are said
to be {\em spatial neighbors} if $x_0 = x'_0$ and for all $i\geq 1$,
$|x'_i - x_i| \leq 1$.

Thus, a {\em discrete geodesics} in $M$ can be defined as a sequence of
points in $M$, $(E_i)_i$ such that for any $i$, the deviation
$w(E_{i-1}, E_i, E_{i+1})$ is a local minimum with respect to spatial
local variations of $E_{i+1}$, that is for any spatial neighbor $G$ of
$E_{i+1}$, we have
\begin{equation}
w(E_{i-1}, E_i, G) \geq w(E_{i-1}, E_i, E_{i+1}) \label{eq:thirdpoint}
\end{equation}
Notice how this condition may be understood as a discrete counterpart
of the Euler-Lagrange equation, in the spirit of
\cite{marsden2001discrete}.

\section{An algorithm to compute a geodesic}\label{sec:algo}

We now give a gradient descent-like algorithm to compute a discrete
geodesic, $\langle t_0, A_0 \rangle, \langle t_1, A_1 \rangle, \langle
t_2, A_2 \rangle$ given a deviation function $w$, a timeline $t_0,
t_1, t_2, ...$, and two starting points $A_0$ and $A_1$.

Assume, $A_{i-1}$ and $A_{i}$ are computed. To compute $A_{i+1}$ start
with a point $\langle t_{i+1}, C\rangle$. Compute $w(\langle t_{i-1},
A_{i-1} \rangle, \langle t_i, A_i \rangle, \langle t_{i+1}, C'
\rangle)$ for all $3^n$ spatial neighbour $C'$ of $C$.  If they are
all larger than $w(\langle t_{i-1}, A_{i-1} \rangle, \langle t_i, A_i
\rangle, \langle t_{i+1}, C \rangle)$ take $C$ for
$A_{i+1}$. Otherwise chose a $C'$ which minimizes $w(\langle t_{i-1},
A_{i-1} \rangle, \langle t_i, A_i \rangle, \langle t_{i+1}, C'
\rangle)$ and iterate, starting from this $C'$.

Whether this iteration will eventually end depends, in general, on
$w(.,.,.)$. For instance, say that $w(\langle t_{i-1}, A_{i-1}
\rangle, \langle t_i, A_i \rangle, \langle t_{i+1}, C' \rangle)$
increases as soon as $A_iC'>t_{i+1}-t_i$. Then $A_{i+1}$ will have
to lie within distance $t_{i+1}-t_i$ of $A_i$, thereby imposing a
bounded velocity $c = 1$, as well as enforcing termination.

\section{Distance induced deviation function}\label{sec:distance}

Most of the times, the idea of deviating from going straight is
induced from a notion of distance. Here is how. Suppose a distance
function $d$, and define the three point distance function
$$l(E,F,G) = d(E,F) + d(F,G).$$
Intuitively, ${\bf FG}$ is understood to deviate from ${\bf EF}$ if it ``leans'' in some spatial direction ${\bf FF'}$, as witnessed by the fact that 
$$l(E,F',G) < l(E,F,G)$$
for $F'$ some neighbour of $F$.

In a continuous-space discrete-time setting this would be formalized
by letting $w(E,\langle x_0, \ldots, x_{n}\rangle,G)$ be
\begin{align}
\partial_0 l (E,\langle x_0, \ldots, x_{n} \rangle,G))^2 + 
\ldots + (\partial_n l (E,\langle  x_0, \ldots, x_{n} \rangle,G))^2.\label{eq:contw}
\end{align}
Thus, for continuous-space discrete-time geodesics $(E_i)_{i}$ each
point $E_i$ is a local extremum for $l(E_{i-1},E_i, E_{i+1})$.

In the discrete spacetime case, $w(E,\langle x_0, \ldots,
x_{n}\rangle,G)$ is simply  \linebreak obtained
by replacing partial derivatives
with finite differences in Equation \eqref{eq:contw}:\linebreak
$\partial_\mu l (E,\langle  x_0, \ldots, x_{n} \rangle,G)$
becomes
\begin{align}
(l(E,\langle  x_0, \ldots,x_\mu -1,\ldots, x_{n} \rangle,G) - 
l(E,\langle  x_0, \ldots, x_\mu +1,\ldots x_{n} \rangle,G))/2.\label{eq:discretew}
\end{align}
And, for discrete spacetime geodesics $(E_i)_{i}$ each point $E_i$ minimizes the possibly non-zero $w(E_{i-1}, E_{i}, E_{i+1})$.

 \section{Discrete Schwarzschild spacetime}\label{sec:planet}

In this section, we give an example of discrete spacetime, which is a
discretization of the Schwarzschild spacetime of General Relativity.

Discretize spacetime down to $\Delta = 1cm$.  Consider a star of
mass $M=2.10^{30}kg$---alike the Sun. Its Schwarzschild 
radius is $m =
2 {\cal G} M / c^2 = 3km = 3.10^5 cm$.  In order to evaluate
distances, consider the metric tensor
\begin{align*}
g(\langle t, x, y \rangle) &=\left(\begin{array}{ccc}
1- \frac{m}{r}&0&0\\
0&-\frac{x^2}{r(r-m)}-\frac{y^2}{r^2}&-\frac{mxy}{r^2(r-m)}\\
0&-\frac{mxy}{r^2(r-m)}&-\frac{x^2}{r^2}-\frac{y^2}{r(r-m)}
\end{array}\right)
\end{align*}
where $r = \sqrt{x^2+y^2}$, and let the distance function $d$ be defined by 
$$d(E,F) = \sqrt{{\bf EF}^{\dagger} g(E) {\bf EF}}.$$ We study the
geodesics trajectory of a planet, with respect to a timeline $0$,
$\tau$, $2\tau$, $3\tau$, ... with 
$a = 10^7$ and $\tau=a\Delta$. Thus $\tau = 10^7 cm = 3.33.10^{-4} s$.
The fake planet has parameters
chosen so as to maximize relativistic effects: its first point is
$E= \langle x_0=0, x_1=10^8 cm=1000km, x_2=0 \rangle$, and its
initial velocity is $vx = 0, vy = 2.10^{-2} c = 6000 km.s^{-1}$.

We compute the geodesics with respect to the $w(.,.,.)$ induced by
$d(.,.)$ as in Section \ref{sec:distance}, and following the algorithm
of Section \ref{sec:algo}. Recall that in this algorithm at iteration
$i$ the point $A_{i+1}$ is found by gradient descent starting from
some point $C$. In the context of planetary movement, a good guess for
$C$ is obtained as follows. Define velocity $S_i=A_i-A_{i-1}$ and
acceleration $R_i=S_i-S_{i-1}$, and make the guess that acceleration
will remain constant, that is $R_{i+1}=R_{i}$. This would entail
that $A_{i+1}=A_i+S_i+R_i$, thus take $C=A_i+S_i+R_i$ as the first
guess and start exploring for the real $A_{i+1}$.  Within reasonable
ranges other heuristics---for instance, $C=A_i+S_i$---lead to the same
trajectories, but may require longer computation times.
 
\begin{figure}[t]
\begin{center}
\includegraphics[scale=1.0]{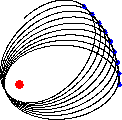}
\end{center}
\caption{{\bf The computed trajectory of a planet.} \label{fig:simulation}}
\end{figure}

A run of the simulation is shown in Figures  \ref{fig:simulation} and
\ref{fig:simulation2}.  Computation time is a few seconds. The code is
available in \cite{ArrighiGeodesics}. The code is easily augmented to
detect aphelion, typically $a=1000 km$, and perihelion, typically $p =
150 km$.

\begin{figure}[t]
\begin{center}
\begin{framed}
{\small 
\begin{center}
\begin{minipage}{8cm}
\begin{verbatim}
Perihelion
t = 3580000000 cm x = -15031004 cm y = -1398397 cm
angle = -174.6849818385271 deg
distance = 1.5095913202506995E7 cm
velocity = 0.13034045668302685 c
---------------------------------------------
Aphelion
t = 7150000000 cm x = 99552205 cm y = 11035292 cm
angle = 6.325378791430483 deg
distance = 1.0016196478647615E8 cm
velocity = 0.019968511905497616 c
theoretical shift = 6.174380835177214 deg
observed shift = 6.210787288401395 deg
---------------------------------------------
Perihelion
t = 10730000000 cm x = -14847048 cm y = -2698767 cm
angle = -169.69789787187233 deg
distance = 1.5090333913952766E7 cm
velocity = 0.13035029101632262 c
---------------------------------------------
Aphelion
t = 14310000000 cm x = 97909925 cm y = 21872062 cm
angle = 12.592542500595265 deg
distance = 1.0032318032058926E8 cm
velocity = 0.019937902185786747 c
theoretical shift = 6.175065136027821 deg
observed shift = 6.267163709164782 deg
\end{verbatim}
\end{minipage}
\end{center}}
\end{framed}
\end{center}
\caption{{\bf Numerics of the computed trajectory of a planet.} 
\label{fig:simulation2}}
\end{figure}

The perihelion shift is visible on Figure \ref{fig:simulation}.
A well-known formula \cite{d1899introducing} states that perihelion
shift in radians per revolution should be 
$$\sigma =  \frac{24 \pi^3 L^2}{T^2 c^2 (1-e^2)} = 
\frac{6 \pi {\cal G} M}{c^2 L (1 - e^2)} = \frac{3 \pi m}{P}$$
where $T$ is the revolution period of the planet, 
$L$ is the semi-major axis of the trajectory of the planet, 
$e$ its eccentricity, and
$P = L (1 - e^2)$ its parameter---recall that, by Kepler's third law,
$T^2 = 4 \pi^2 L^3/({\cal G}M)$, and 
$m = 2 {\cal G} M / c^2$.
Then an easy geometrical relation is
$$P = \frac{2}{1/a + 1/p}$$
hence
$$\sigma = (3/2) \pi m (1/a+ 1/p)$$
which typically is $6.17$ deg. The observed shift is around $6.27$ deg.

\section{Cellular Automata in Mechanics}\label{sec:CAMechs}

As suggested in the Introduction, one motivation for discretizing
General Relativity is to describe the motion of a planet in a cellular
automaton.

Recall that, in the cellular automata vocabulary, a configuration
$\sigma$ is a function which associates, to each cell $C$ of the grid
${\mathbb Z}^n$, some internal state $\sigma(C)$ taken in the set
$\Sigma$. A cellular automaton is a function $F$ from configurations
to configurations, which has the following physics-inspired
symmetries:
\begin{itemize}
\item bounded velocity of propagation of information;
\item homogeneity time and space;
\item bounded density of information, that is $\Sigma$ is finite.
\end{itemize} 

The state of a cell can be used to express the presence or the absence
of a particle in this region of space. This way cellular automata can
describe particle motions. For instance the simplest $n$-dimensional
cellular automata---2 states, radius 1---can describe one particle
motion among $3^n$, as in each dimension, it could have velocity $-1$,
$0$, or $1$.

To describe more complex motions, we must increase the number of
states. For instance in a 1-dimensional automaton with radius 1, we
can describe the motion of a particle that goes to the right at
velocity $1/2$, by alternating states $s_1$---stay still---and $s_2$---step---,
but also the motion of a particle that goes to the right at
velocity $1$, staying in the state $s_3$.

Another option is to increase the radius of the automaton. For
instance, in a 1-dimensional automaton, with radius $2$ we can
describe the motion of a particle that goes on the right at velocity
$1$ staying in a state $s_1$ or at velocity $2$ staying in a state
$s_2$.  Notice that modulo changing the units, the former behaviour
can be obtained from the latter just by cell grouping.

We want to address the following question: to what extent is the
algorithm of Section \ref{sec:algo} just a cellular automaton?

\section{Geodesics as Cellular Automata}\label{sec:CAGeodesics}

Whether the algorithm of Section \ref{sec:algo} enforces a bounded
velocity of propagation of information $c = 1$ depends, in general, on
the properties of $w(.,.,.)$. If such a velocity bound is enforced,
then the motion of the body can be described in a cellular automaton
of radius $r$. It is well-known that the velocity of a particle in a
continuous Schwarzschild spacetime is bounded by $c = 1$. We
conjecture that this is also the case for the discretized
Schwarztchild spacetime.

If $w(.,.,.)$ does not depend on space and time, then the algorithm
clearly acts the same everywhere and everywhen, so that homogeneity is
also enforced. In the important case where it depends upon a
space-dependent metric, then this metric field has to be carried by
the internal state of the cells, even if it does not contain a
particle, so that homogeneity is still enforced.

Let us evaluate whether bounded density of information holds. Even
when $w(.,.,.)$ does not depend on space and time, it is still the
case that if a particle is at $E_i$, we need its velocity ${\bf
E_{i-1}E_i}$ to compute its next position $E_{i+1}$. But thanks to
bounded velocity of propagation, and the fact that positions are
discrete, the number of possible velocities is bounded above by
$b=(2a+1)^n$, so that bounded density of information is preserved. In
the important case of a space-dependent metric carried by the internal
state of the cells, whether bounded density holds depends upon whether
we can assume that the metric field can be given with bounded
precision. Even if this is not the case, notice that for a given cell,
all that matters is to distinguish, for each input velocity of the
particle, between $b$ output candidate target cells. This map is a
discrete counterpart to the connection associated to the metric. It
contains just the finite amount of information that needs to be
attached to the cell in order to compute geodesics. It could in fact
be pre-compiled into each cell, thereby yielding a cellular automaton
with $b+1$ internal states to code for presence and velocity, times
$b^b$ to code for the discrete connection.

\section{Time versus space, precision}\label{sec:precision}

Geodesics have been popularized by General Relativity. General
Relativity likes to put space and time on an equal footing. Numerical
schemes for General Relativity ought to pursue that path, in
particular it would be nice if the timeline of the computed geodesic 
were just $0$, $\Delta$, $2 \Delta$, $3 \Delta$, ...
that is if $a$ was equal to $1$. In
the scheme of Section \ref{sec:algo}, this choice leads to a cellular
automaton of radius $1$, which is appealing, but it also restricts to
$b=3^n$ the number of possible velocities. As we discussed in Section 
\ref{sec:CAMechs}, 
this
severely limits the number of motions that can be described. In the
quantum setting, superpositions of basic velocities may compensate for
this \cite{di2013quantum,di2014quantum,ArrighiGRDirac}. Classically,
this is dramatic loss in precision. This is why, 
in Section \ref{sec:planet}, we took $a = 10^7$.

However, we also saw that a radius of $a'=1$ can be obtained from a
cellular automaton of arbitrary radius $a$ simply by grouping each
$a^n$ hypercube of cells into one supercell. Each supercell now has an
internal state in $\Sigma'=\Sigma^{a^n}$. Notice that keeping the
position of the single particle within the hypercube is
crucial. Otherwise, all the velocities of norm less than one supercell
are rounded up to the center of the supercell---and so the increased
precision in the velocities is not much use. Hence, $\Sigma'$ is
really just coding for a velocity amongst $b$ possibilities, which is
appealing\ldots\ but also for the position of the single particle
within the hypercube, which perhaps is not so satisfactory. After all,
what this space grouping has done is really just to hide the
discrepancy between the disctetization step $\Delta$ and the 
computed geodesics timeline step $a \Delta$, by hiding some of
spatial precision within the internal space of the supercells.

Hence, $a \gg 1$ appears to be fundamental requirement for
precision. Notice that large values for $a$ are better obtained by
diminishing the discretization step $\Delta$ rather than augmenting
the timeline step $a \Delta$, as we cannot hope to achieve a
pseudo-elliptic trajectory with just a handful of velocity changes per
revolution. Running the simulations, it was indeed observed that $a$
large, for instance $a=10^7$, yields increased stability.  But only to
some extent: after a while the number of possible velocities
$b=(2a+1)^n$ exceeds those which can be stored as a vector of
machine-sized integers. 

It also helps to fine-grain the discretization step $\Delta$, keeping
$a$ constant. Running the simulations, it was indeed observed that
this yields increased stability and convergence---at the expense of
(reasonably) longer computation times. At some point, however, the
finite-differences of \eqref{eq:discretew} can become unstable, due to
very small differences between $l(E,F,G)$ and $l(E,F',G)$ when
$FF'=1$, again hitting bounded machine floating point-arithmetic
precision---but this can easily be fixed by evaluating these
derivatives with $FF'$ a fraction of $l(E,F,G)$ independent of
$\Delta$.

\section{Recovering continuous spacetime geodesics}\label{sec:continuum}

The algorithm of Section \ref{sec:algo} is successful in computing
geodesics in discrete time and space ${\mathbb Z} \times {\mathbb
  Z}^{n}$, in a way which is consistent with continuous-space
discrete-time geodesics. We now explain how continuous-space
discrete-time geodesics are themselves consistent with the standard
geodesics of the fully continuous setting. For this question to make
sense, we place ourselves in the case of Section \ref{sec:distance}: a
distance-induced deviation function.

\begin{figure}\centering
\begin{tikzpicture}
\tikzstyle{every node} = [label distance=1pt, circle, fill, inner sep=0pt, minimum size=5pt]
\tikzset{>=latex}
\node[label=45:$E$] (E) at (1,0) {};
\node[label=45:$F$] (F) at (0,2) {};
\node[label=180:{\color{gray}$F'$},gray] (F') at (-1.75,2) {};
\node[label=45:$G$] (G) at (-2,4) {};
\draw[->] (E) -- (F) node [midway,fill=none,label=45:$\varepsilon v$,label=-93:${g(E)}$] {};
\draw[->] (F) -- (G) node [midway,fill=none,label=45:$\varepsilon v'$,label=-95:${g(F)}$] {};
\draw[->,gray,dashed] (F) -- (F') node [midway,fill=none,label=-45:$\delta d$] {};
\draw[->,gray] (E) -- (F') node [midway,fill=none] {};
\draw[->,gray] (F') -- (G) node [midway,fill=none] {};
\end{tikzpicture}
\caption{{\bf The continuum limit} is obtained for $\varepsilon$ and $\delta$ tending to zero.\label{fig:limit}}
\end{figure}
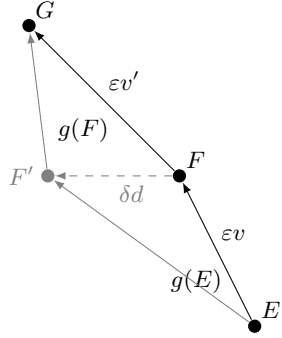

As in Figure \ref{fig:limit}, consider three points $E$, $F$, $G$, taken at successive times $0, \tau, 2\tau$.
Let $\varepsilon$ be the distance
$EF$, measured according to $g(E)$, that is the proper time along
${\bf EF}$ and $v = {\bf EF}/\varepsilon$. 
In the same way, let $\varepsilon'$ be the distance $FG$ according to 
$g(F)$ and $v' = {\bf FG}/\varepsilon'$. 

We said that trajectory
$EFG$ is a continuous-space discrete-time geodesics if and only if it
minimizes the distance $EF+FG$, with respect to infinitesimal changes
of $F$ into $F'$. Let us take 
$FF'=\delta d$ where $d$ is a vector, normal with respect to $g(F)$.

The distance $EF'+F'G$ is given by:
$$\sqrt{(\varepsilon v+\delta d)^\dagger g(E) (\varepsilon v+\delta d) }+\sqrt{(\varepsilon v'-\delta d)^\dagger g(F') (\varepsilon v'-\delta d)}$$
Consider the first term. Its derivative with respect to $\delta$ is 
$$\frac{d^\dagger g(E) (\varepsilon v+\delta d)+(\varepsilon v+\delta
d)^\dagger g(E) d}{2\sqrt{(\varepsilon v+\delta d)^\dagger g(E)
(\varepsilon v+\delta d)}}$$
Taken at $\delta=0$ and using the symmetry of $g(E)$ we get:
$$\frac{d^\dagger g(E) (\varepsilon v)}{\sqrt{(\varepsilon v)^\dagger g(E) (\varepsilon v) }}=d^\dagger g(E) v$$
Consider the second term. If $g(F')$ were just $g(F)$, the same
process would yield $-d^\dagger g(F) v'$. We would then just have
$d^\dagger g(E) v-d^\dagger g(F) v'=0$, yielding $v'=g(F)^{-1}g(E)
v$. This is the equation derived in \cite{williams1981regge}, and is
in the same spirit as that obtained \cite{polthier2006straightest} in
the framework of simplicial complexes. Unfortunately it does not yield
accurate predictions for perihelion shift, as pointed out in
\cite{williams1981regge,brewin1993particle} and confirmed by our
simulations. This is because one has to take into account that the
variations of $g(F)$ around $F$ yield a third term:
$$\frac{(\varepsilon' v')^\dagger(\partial g(F) .d) 
(\varepsilon' v')}{2\sqrt{(\varepsilon' v')^\dagger g(F) 
(\varepsilon' v') }}=\frac{\varepsilon'}{2}v'^\dagger(\partial g(F) .d)v'$$
Let us emphasize that straightest geodesics on simplicial complexes
\cite{polthier2006straightest} do not see this term either: quite
simply because a path $EFG$ between two adjacent simplices sees the
geometry of the first simplex---that is the term $g(E)$---and the geometry
of the second simplex---that is the term $g(F)$---, but ignores the
variations of the geometry in some arbitrary direction $FF'$. In other
words, simplices are usually thought of as polyhedrons of constant
metric---but in order to be consistent with the continuum they must
be interpreted as surfaces of constant metric derivatives. Altogether,
we get that trajectory $EFG$ is a geodesics if and only if
\begin{align}
 d^\dagger \left(g(F) v'-g(E) v\right)=\frac{\varepsilon'}{2} v'^\dagger(\partial g(F) .d)v'
\end{align}
in every directions $d$. 
Unfortunately, this is still inconvenient to solve for $v'$. At this
stage the traditional, continuous approach to geodesics follows two
simplifying steps, which in the discrete setting translate into two
approximations. The first step is to evaluate the condition for $d$
only along the coordinate directions:
\begin{align}
\left(g(F)_{\lambda.} v'-g(E)_{\lambda.} v\right)=\frac{\varepsilon'}{2} v'^\dagger(g(F)_{,\lambda})v'
\end{align}
for all $\lambda$. 
The second step is to realise that ${\bf FG}={\bf EF}+O(\varepsilon^2)$, and hence that $v'=v+O(\varepsilon)$ and $\varepsilon'=\varepsilon+O(\varepsilon^2)$.  
Indeed, in the continuum we assume that the time trajectory $x(t)$ is twice diffentiable hence is has a first order expansion in $\tau$, from which we get ${\bf FG}={\bf EF}+O(\tau^2)$ and $EF\sim \tau\frac{dx}{dt}$, i.e.  $\varepsilon \sim \tau$. 

We get:
\begin{align}
g(F)_{\lambda.} v'&=g(E)_{\lambda.} v+\frac{\varepsilon}{2} v^\dagger(g(F)_{,\lambda})v.\\
 v'&= {g(F)}^{-1}_{\cdot\lambda}\left(g(E)_{\lambda\cdot} v + \frac{\varepsilon}{2} v^\dagger g(F)_{,\lambda} v \right)
\end{align}
We get:
\begin{align}
g_{\lambda\cdot}\,v'&= g_{\lambda\cdot}\,v -\varepsilon g_{\lambda\cdot,\mu}v^\mu v + \frac{\varepsilon}{2} v^\dagger g_{,\lambda} v\nonumber\\
g_{\lambda\cdot}\,(v'-v)&= \frac{\varepsilon}{2} \left( v^\dagger g_{,\lambda} v- 2g_{\lambda\nu,\mu}v^\mu v^\nu\right)\nonumber\\
g_{\lambda\cdot}\,(v'-v)&= \frac{\varepsilon}{2} \left(g_{\mu\nu,\lambda}v^\mu v^\nu- g_{\lambda\nu,\mu}v^\mu v^\nu - g_{\lambda\mu,\nu}v^\mu v^\nu\right)\nonumber\\
(v'-v)&=-\varepsilon\Gamma_{\mu\nu}v^\mu v^\nu\label{eq:T1}
\end{align}
where
\begin{align}
\Gamma_{\mu\nu}=g^{-1}_{\cdot\lambda}\left(g_{\lambda\nu,\mu} + g_{\lambda\mu,\nu} - g_{\mu\nu,\lambda} \right)/2\label{eq:T2}
\end{align}
Let us study the continuum limit $\varepsilon \to 0$. Suppose $E$, $F$, $G$ were three consecutive points along a curve $x$ parametrized by its proper-time $s$. Since $\varepsilon$ was the proper time along $EF$, 
we have
$$\frac{dx}{ds} = \lim_{\varepsilon \to 0} {\bf EF}/\varepsilon = 
\lim_{\varepsilon \to 0} v$$
and 
$$\frac{d^2x}{ds^2} = \lim_{\varepsilon\to 0} 
({\bf FG} - {\bf EF})/\varepsilon^2 = 
\lim_{\varepsilon\to 0} (v' - v)/\varepsilon$$
and hence
\begin{align}
\frac{d^2 x}{ds^2} =-\Gamma_{\mu\nu}\frac{dx^\mu}{ds} \frac{dx^\nu}{ds} \label{eq:T3}
\end{align}
which is your traditional geodesics equation. 

\section{Conclusion}\label{sec:conclusion}

Summarizing, we have introduced a generic notion of discrete geodesics
as straightest trajectories in discretized spacetime, which are such
that any three successive points $E, F, G$ must minimize the deviation
function $w(E,F,G)$. Given $E$ and $F$, $G$ is implicitly determined:
this can be viewed as a dynamical system and computed via a gradient
descent algorithm. For a metric space, the canonical choice for
$w(.,.,.)$ measures how the length $EF$+$FG$ varies with small
variation of $F$. This was validated numerically, by computing the
trajectory of a planet in discretized Schwarzschild spacetime, and
recovering a perihelion shift of the right order. This was also
validated by taking the continuum limit and recovering the standard
geodesics equations on pseudo-riemannian manifolds.

Part of our motivations were to evaluate the strength and limits of
cellular automata. Recall that three well-accepted postulates about
physics---bounded velocity of propagation of information, homogeneity in
time and space, and bounded density of information---necessarily imply
that physics may be cast in the framework of cellular automata---both
in the classical and quantum settings \cite{Gandy,ArrighiGandy}.  Both
theorems, however, rely on the implicit hypothesis of a flat
spacetime. To which extent can cellular automata account for
relativistic trajectories, that is geodesics? This paper shows that
discrete geodesics can be cast in the framework of cellular automata,
provided that a few extra assumptions are met: that the metric can be
given with bounded precision, and that it has the property of fixing a
velocity limit. These extra assumptions do not contradict the three
postulates: they are but instances of them.

Yet, this paper shows that a large discrepancy between the time
discretization step the space discretization step is necessary in
order to maintain a good precision on the velocity of
particles. Namely, the number of particle velocities varies in
$(2a+1)^n$, with $n$ the dimension of space and the radius $a$ of the
cellular automaton, which is therefore inherently large. Thus,
computing discrete geodesics---and straight lines in euclidean space,
for that matter---is local\ldots but not that local. This may come as a
surprise, and suggest that geodesics equations are better-behaved in the
continuum. An alternative is to live with imprecise velocities. A
planet is a collection of particles, and so it may be the average of
their imprecise velocities which grants it a precise averaged
velocity. In fact, a single particle is itself quantum, and may thus
be in a superposition of these imprecise velocities, yielding a
precise averaged velocity---as is made formal in the eikonal
approximation \cite{cianfrani2008dirac}. This is in fact precisely
what happens in quantum cellular automata models of quantum particles
in curved spacetime, as shown in
\cite{di2013quantum,di2014quantum,ArrighiGRDirac}. All of these
considerations suggest that nature's way of working out timelike
geodesics trajectories may in fact be emergent, from the simpler and
more local behaviour of spinning particles.

Hence, for future work, it may be interesting to look for discrete
models based on spinning particles, oscillating along a few, cardinal,
light-like directions. These may in fact be closer to mimicking the
real behaviour of fermions in curved spacetime, with the hope to
recover the Mathisson-Papapetrou-Dixon equation---a generalization of
the geodesics equation to spatially extended massive spinning bodies---as 
emergent, in analogy with the continuum
\cite{cianfrani2008dirac}. Such discrete models may be more
local. Another approach is to work directly in terms of a discrete
connection \cite{lorenzi2011schild}. In the continuum, the Levi-Civita
connection is axiomatized as being the unique metric-compatible and
torsion-free connection. That given in Equation \eqref{eq:T2} is exactly
torsion-free, but interpreted as a discrete connection, as in
Equation \eqref{eq:T1}, it is metric-compatible only to first order. One
can ask for both properties to be met exactly even in the discrete
setting, this specifies the intersection of two ellipses. In
$2$-dimensions the number of solutions is finite, but this is not even
the case in higher-dimensions: the axiomatization suggested by the
continuum breaks down and demands fixing.

\section*{Acknowledgements} 

This work has been funded by the ANR-12-BS02-007-01 TARMAC grant, the
ANR-10-JCJC-0208 CausaQ grant, and the John Templeton Foundation,
grant ID 15619. Pablo Arrighi benefited from a visitor status at the
IXXI institute of Lyon.

\bibliography{perihelion}

\bibliographystyle{plain}

\end{document}